\documentclass[conference]{IEEEtran}
\IEEEoverridecommandlockouts
\usepackage{cite}
\usepackage{amsmath,amssymb,amsfonts}
\usepackage{algorithmic,booktabs,multirow,subfigure}
\usepackage{graphicx}
\usepackage{textcomp, hyperref}
\usepackage{xcolor}
\def\BibTeX{{\rm B\kern-.05em{\sc i\kern-.025em b}\kern-.08em
T\kern-.1667em\lower.7ex\hbox{E}\kern-.125emX}}
\begin{document}

\title{SVM and ANN based Classification of EMG signals by using PCA and LDA \\
{\footnotesize Influenced by: Toward improved control of prosthetic fingers using surface electromyogram (EMG) signals \cite{b1}}
}

\makeatletter
\newcommand{\linebreakand}{%
\end{@IEEEauthorhalign}
\hfill\mbox{}\par
\mbox{}\hfill\begin{@IEEEauthorhalign}
}
\makeatother
\author{\IEEEauthorblockN{Hritam Basak}
\IEEEauthorblockA{\textit{Roll: 001710801018} \\
\textit{BEE-IV}\\
\textit{Department of Electrical Engineering}\\
Jadavpur University }
\and
\IEEEauthorblockN{Alik Roy}
\IEEEauthorblockA{\textit{Roll : 001710801038} \\
\textit{BEE-IV}\\
\textit{Department of Electrical Engineering}\\
Jadavpur University }
\and
\IEEEauthorblockN{Jeet Bandhu Lahiri}
\IEEEauthorblockA{\textit{Roll : 001710801040} \\
\textit{BEE-IV}\\
\textit{Department of Electrical Engineering}\\
Jadavpur University }
\linebreakand
\IEEEauthorblockN{Sayantan Bose}
\IEEEauthorblockA{\textit{Roll : 001710801078} \\
\textit{BEE-IV}\\
\textit{Department of Electrical Engineering}\\
Jadavpur University }
\and
\IEEEauthorblockN{Soumyadeep Patra}
\IEEEauthorblockA{\textit{Roll : 001710801082} \\
\textit{BEE-IV}\\
\textit{Department of Electrical Engineering}\\
Jadavpur University }

}

\maketitle

\section{Introduction}
In recent decades, biomedical signals have been used for communication in Human-Computer Interfaces (HCI) for medical applications; an instance of these signals are the myoelectric signals (MES), which are generated in the muscles of the human body as unidimensional patterns. Because of this, the methods and algorithms developed for pattern recognition in signals can be applied for their analyses once these signals have been sampled and turned into electromyographic (EMG) signals. Additionally, in recent years, many researchers have dedicated their efforts to studying prosthetic control utilizing EMG signal classification, that is, by logging a set of MES in a proper range of frequencies to classify the corresponding EMG signals.
The EMG signals are obtained from sensors placed on the skin surface and can help retrieve muscular information during contractions when flexing or extending an articulation. There are also implants placed under the skin that facilitate signal acquisition, but these are not commonly used. Regarding the pattern recognition problem for myoelectric control systems, its success depends mostly on the classification accuracy because myoelectric control algorithms are capable of detecting movement intention; therefore, they are mainly used to actuate prostheses for amputees.
To carry out the pattern recognition for myoelectric applications, a series of features is extracted from the myoelectric signal for classification purposes. The feature classification can be carried out on the time domain or by using other domains such as the frequency domain (also known as the spectral domain), time scale, and time-frequency, amongst others.
One of the main methods used for pattern recognition in myoelectric signals is the Support Vector Machines (SVM) technique whose primary function is to identify an n-dimensional hyperplane to separate a set of input feature points into different classes. This technique has the potential to recognize complex patterns and on several occasions, it has proven its worth when compared to other classifiers such as Artificial Neural Network (ANN), Linear Discriminant Analysis (LDA), and Principal Component Analysis(PCA). The key concepts underlying the SVM are (a) the hyperplane separator; (b) the kernel function; (c) the optimal separation hyperplane; and (d) a soft margin (hyperplane tolerance).
So, by summarizing, this project has the following three-fold contribution:
\begin{itemize}
\item We have used different feature extraction methods to extract suitable and distinctive features from the dataset for ease of classification.
\item We have used two different data analysis methods named Principal Component Analysis (PCA) and Linear Discriminant Analysis (LDA) for dimensionality reduction of the feature set for the sake of reducing the computational cost.
\item Finally we have used two different classifiers SVM and MLP for the classification of different action classes.
\end{itemize}

\section{Description of the dataset}
EMG signal for hand movement was acquired from the Rami Khushaba repository. The dataset is entitled as \textbf {DETECTING INDIVIDUAL AND COMBINED FINGERS MOVEMENTS}. Raw data of 8 persons and each of them having 10 different classes of individual and combined fingers movements including the flexion of each of the individual fingers, i.e., \textbf{Thumb(T), Index (I), Middle (M), Ring (R), Little (L) and the pinching of combined Thumb–Index (T–I), Thumb–Middle (T–M), Thumb–Ring (T–R), Thumb–Little (T–L)}, and finally the hand close \textbf{(HC)}. Now, every 10 classes have 6 files which contain 20,000 EMG data within 2 EMG channels each. So, we are left with a huge data collection of dimension \textbf{[(8 x 10 x 6) x (20000 x 2)] } where (20,000 x 2) can be featured as row and column of a matrix respectively.

\subsection{Goal of the project}
The goal of this work is to increase the classification accuracy where ten classes of individual and combined finger movements are to be recognized that can be of future help for the prosthetic hand movements.
\subsection{Few Points Regarding Properties of EMG Signals}
The amplitude of the EMG signal is stochastic (random) with a Gaussian distribution that ranges from 0 to 10 mV (peak to peak). Two parameters are commonly used to measure the amplitude:
\begin{enumerate}
\item the root-mean-square (RMS) value
\item the mean absolute (MA) value.

\end{enumerate}
The fidelity of an EMG signal is influenced by two main concerns:
\textbf{Signal to noise ratio} - the ratio of the energy in the EMG signal to the energy in the noise signal.

\textbf{Distortion of the signal} - the relative contribution of any frequency component in the EMG signal should not be altered.

\section{Proposed approach}
For the classification task, we have used the raw data directly to fit the classifier following the dimension reduction, however, that resulted in some poor results and accuracy in the range of 10-20\%, calling for specific feature extraction tasks.
\subsection{Feature extracion}
Following the original work of \cite{b1}, 8 feature sets are extracted from the pre-processed raw EMG signals. So, 8 features are available for 2 channels each which results in 16 features. Now, we calculate first-order moments to seventh order moments for each channel. As a result, we get total 30 features (2 channels × 8 features = 14 features + 7 + 7 moments). Features in the time domain are more commonly used for EMG pattern recognition. This is because they are easy and quick to calculate as they do not require any transformation. Time-domain features are computed based upon the input signals amplitude. The resultant values give a measure of the waveform amplitude, frequency, and duration with some limitations.

\subsubsection{Root mean square (RMS) value}
The RMS value represents the square root of the average power of the EMG signal for a given period. It is a time-domain variable because the amplitude of the signal is measured as a function of time. RMS value of a signal f(t) over the interval T\textsubscript{1} to T\textsubscript{2} is given by:
\begin{equation}
X_{RMS}=\sqrt{\frac{1}{T_2-T_1}\int_{T_1}^{T_2}[f(t)]^2 dt}
\end{equation}
\subsubsection{Mean absolute value}
The mean absolute (MA) value is the computer calculated equivalent of the average rectified value (ARV). The MA value is a time-domain variable because it is measured as a function of time. It represents the area under the rectified EMG signal. The MA value is used as a measure of the amplitude of the EMG signal like the root mean square value. It is given by the following equation over time T:
\begin{equation}
ARV=\int_t^{t+T}|X_i|dt
\end{equation}

where T is the period over which the signal is integrated.
However, the RMS value is often preferred over the MA value because it provides a measure of the power of the EMG signal while the MA value does not.
\subsubsection{Zero crossing}
Zero crossing (ZC) is the number of times that the amplitude value of EMG signal crosses the y = 0 line. In EMG feature, the threshold condition is used to abstain from the background noise. This feature provides an approximate estimation of frequency domain properties. It can be formulated as:
\begin{equation}
ZC=\sum_{n=1}^{N-1}[sgn(X_n-X_{n+1})\cap|X_n-X_{n+1}| ]>Threshold
\end{equation}
\subsubsection{Slope sign change}
Slope Sign Change (SSC) is similar to ZC. It is another method to represent the frequency information of EMG signal. The number of changes between positive and negative slope among three consecutive segments are performed with the threshold function for avoiding the interference in EMG signal. The calculation is defined as:
\begin{equation}
SSC=\sum_{n=2}^{N-1}[f(X_n-X_{n+1})\times(X_n-X_{n-1})]
\end{equation}
\subsubsection{Autoregressive features}
Autoregressive (AR) model described each sample of EMG signal as a linear combination of previous samples plus a white noise error term. AR coefficients are used as features in EMG pattern recognition. The model is basically of the following form:
\begin{equation}
x_n=\left[ -\sum_{i=1}^p a_iX_{n-i}+W_n \right]
\end{equation}

where x\textsubscript{n} is a sample of the model signal, a\textsubscript{i} is AR coefficients, w\textsubscript{n} is white noise or error sequence, and p is the order of the AR model.

\subsubsection{Sample skewness}
The skewness of a random variable X is the third standardized moment M\textsubscript{3} defines as:
\begin{equation}
S_k=E\left[ \left( \frac{X-\bar{X}}{\sigma} \right)^3 \right]=\frac{M_3}{\sigma ^3}
\end{equation}
Where $\sigma$ is the standard deviation and M\textsubscript{3} is the third moment.

\begin{figure*}[t]
\centering
\includegraphics[width=1.8\columnwidth]{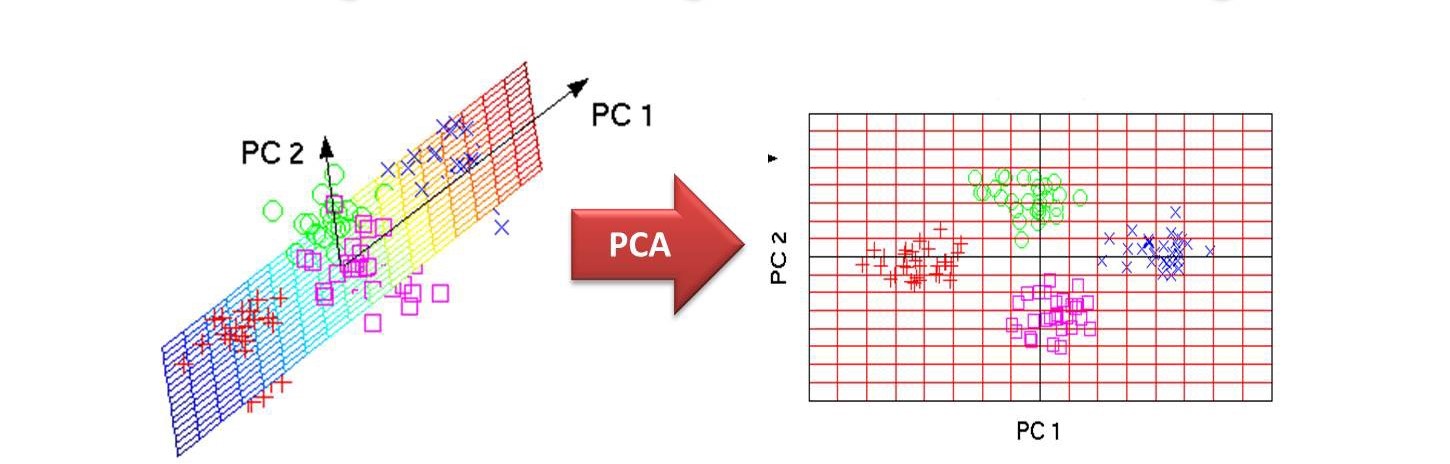}\caption{Dimensionality reduction by using PCA}
\end{figure*}
\subsubsection{Waveform length}
Waveform length (WL) is the cumulative length of the waveform over the time segment. WL is related to the waveform amplitude, frequency and time. It is given by
\begin{equation}
Wl=\sum_{n=2}^{N-1}|X_n-X_{n+1}|
\end{equation}
\subsubsection{Integrated absolute value}
Integrated EMG (IEMG) is calculated as the summation of the absolute values of the EMG signal amplitude. Generally, IEMG is used as an index to detect the muscle activity that used to oncoming the control command of assistive control device. It is related to the EMG signal sequence firing point, which can be expressed as:
\begin{equation}
IAV=\sum_{i=1}^n|X_i|
\end{equation}
\subsubsection{Moments}
The moments of a function are quantitative measures related to the shape of the function's graph. The equation of the nth moment is given below:
\begin{equation}
M_x=E[X-\bar{X}^n]
\end{equation}

\subsection{Dimensionality reduction}

We have reduced the dimension of the feature matrix for the sake of computation. PCA and LDA are two popularly known methods for dimensionality reduction. Hence we have used these methods for this purpose.
\subsubsection{\textbf{Principal Component Analysis (PCA)}}
Principal component analysis (PCA) is a multivariate technique that analyzes a data table in which observations are described by several inter-correlated quantitative dependent variables. Its goal is to extract the important information from the statistical data to represent it as a set of new orthogonal variables called principal components, and to display the pattern of similarity between the observations and of the variables as points in spot maps. So, PCA uses an orthogonal transformation to convert a set of variables that may or may not be correlated into a set of uncorrelated variables. As a result, this method reduces the dimensionality of a large data set by transforming a large set of variables into a smaller one without the loss of such information.

The process of obtaining principal components from a large raw dataset is explained in the following steps:
\begin{enumerate}
\item Organize a data set as an m × n matrix, where m is the number of measurement types and n is the number of trials.
\item The mean of the entire data should be calculated for each of the data dimensions in the whole data set. Then we have to normalize the data so that PCA works properly. This is done by subtracting the respective means from the numbers in the respective column.
\item Then we have to calculate the Covariance Matrix corresponding to our raw dataset using the following formula 
\begin{equation}
cov(x,y)=\frac{1}{n-1}\sum_{i=1}^{n} [(x_i-\bar{x})\cdot (y_i-\bar{y})]
\end{equation}
where $\bar{x}$= arithmetic mean of data X, $\bar{y}$= arithmetic mean of data Y, n = number of observations.

\textbf{An example:} We’ll make up the covariance matrix for an imaginary 3-dimensional data set, using the usual dimensions x, y, and z. Then, the covariance matrix has 3 rows and 3 columns, and the values are this:

\begin{center}
$
\begin{bmatrix} 

(cov(x,x)&cov(x,y)&cov(x,z)\\cov(y,x)&cov(y,y)&cov(y,z)\\cov(z,x)&cov(z,y)&cov(z,z))\\
\end{bmatrix}
$
\end{center}

Some points that are taken into consideration is that the diagonal elements are the covariance value between one of the dimensions with itself. These are the variances for that particular dimension. The other point is that the cov(x,y)=cov(y,x) makes the matrix symmetrical about the main diagonal. However, the covariances that we have as the entries of the matrix can also give us the idea of the correlations between the variables. If the sign of the covariance between two variables is positive, then the two variables will increase or decrease i.e. they are correlated and vice versa. 
\item Calculate the Eigenvectors and Eigenvalues of the Covariance Matrix.
\item The next step is to choose the components (i.e. the eigenvectors) and forming the feature vector. In general, once eigenvectors are found from the covariance matrix, the next step is to order them by eigenvalue, highest to lowest. This gives the components in order of significance. After that, the components with lesser significance can be ignored as per the requirement of the researcher. What needs to be done now is we need to form a feature vector, which is just a fancy name for a matrix of vectors. This is constructed by taking the eigenvectors that we want to keep from the list of eigenvectors and forming a matrix with these eigenvectors in the columns.
\item Once the components (eigenvectors) are chosen to keep in the data and a feature vector has been formed, then the transpose of the vector should be calculated and multiplied on the leftover of the scaled original data set, transposed.
\end{enumerate}

So, thus PCA gives us multiple principal components for multidimensional data (PCs $\leq$ Dimension of the data), but it tries to squeeze most of the information within the initial variables into the first components, hen maximum remaining information in the second component, and so on. But the new variables being a mixture of initial variables are combined in such a way that they are uncorrelated. 

Due to its capability of reducing the dimension of a huge dataset without losing much information, PCA is predominantly used in the domain like facial recognition, computer vision, and image compression. It is also used for finding patterns in data of high dimension in the field of finance, data mining, bioinformatics, etc.

\subsubsection{\textbf{Linear Discriminant Analysis (LDA)}}
Linear Discriminant Analysis (LDA) is, like Principle Component Analysis (PCA), a method of dimensionality reduction. However, both are quite different in the approaches they use to reduce dimensionality. While PCA chooses new axes for dimensions such that variance (and hence the ‘shape’) of the data is preserved, LDA chooses new axes such that the separability between two classes is optimized, and hence is a supervised technique.
Hence, when one discusses using dimensionality reduction, not for visualization purposes (which would require retention of shape) but to increase model performance, usually they are talking about LDA. Both, however, represent the newly formed dimensions in terms of linear combinations of dimensions in the dataset. Beyond simply support for model-building, however, it has proven itself a powerful method for analysis and interpretation.

Linear Discriminant Analysis is a three-step process:
\begin{enumerate}
\item Calculate the ‘separability’ between the classes. Known as the between-class variance, it is defined as the distance between the mean of different classes and allows for the algorithm to put a quantitative measure on ‘how difficult’ the problem is (closer means = harder problem). This separability is kept in a ‘between-class scatter matrix’.
\item Compute the within-class variance or the distance between the mean and the sample of every class. This is another factor in the difficulty of separation — higher variance within a class makes the clean separation more difficult.
\item Construct a lower-dimensional space that maximizes the between-class variance (‘separability’) and minimizes the within-class variance. Known as Fisher’s Criterion, Linear Discriminant Analysis can be computed using singular value decomposition, eigenvalues, or using the least-squares method.
\end{enumerate}

\textbf{Example of LDA} For example, consider a very noisy two-dimensional scatterplot with two classes, which is now projected by LDA into one dimension as shown in the figure below.
\begin{figure}[h]
\centering
\includegraphics[width=\columnwidth]{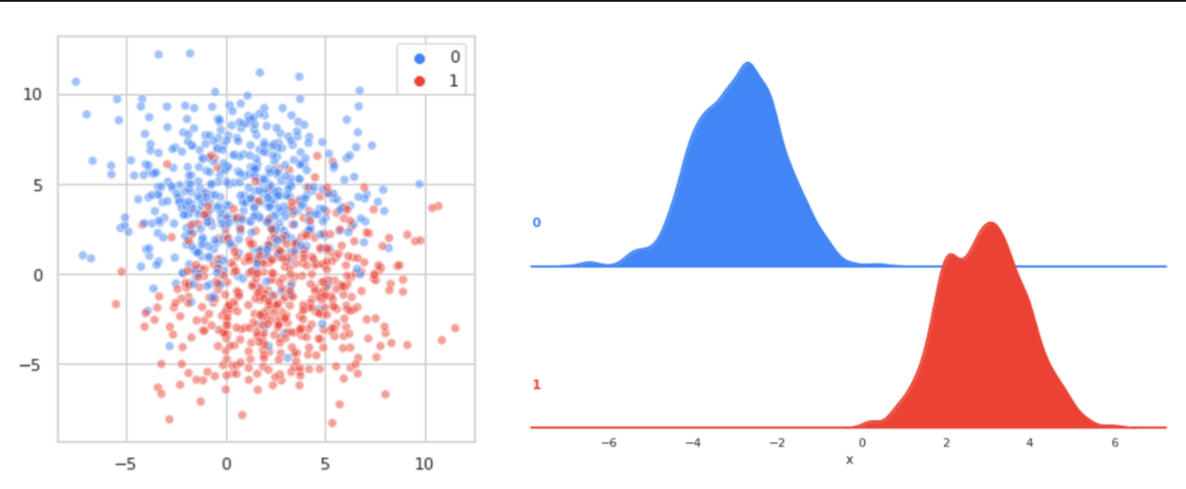}\caption{Graphical representation of Linear Discriminant Analysis (LDA)}
\end{figure}
As seen in the figure, the separation between the two classes is pretty clear and distinct.

So, Linear Discriminant Analysis is not only a dimensionality reduction technique but also it is a simple and effective method for classification and that is why this method is quite popular nowadays in various machine learning and deep learning applications. There are also some popular extensions of LDA like Quadratic Discriminant Analysis (QDA), Flexible Discriminant Analysis (FDA), Regularized Discriminant Analysis (RDA).

\subsection{Classification}
We have used SVM and ANN for our classification purpose.
\subsubsection{\textbf{Support Vector Machine (SVM)}}
\begin{figure}[h]
\centering
\includegraphics[width=\columnwidth]{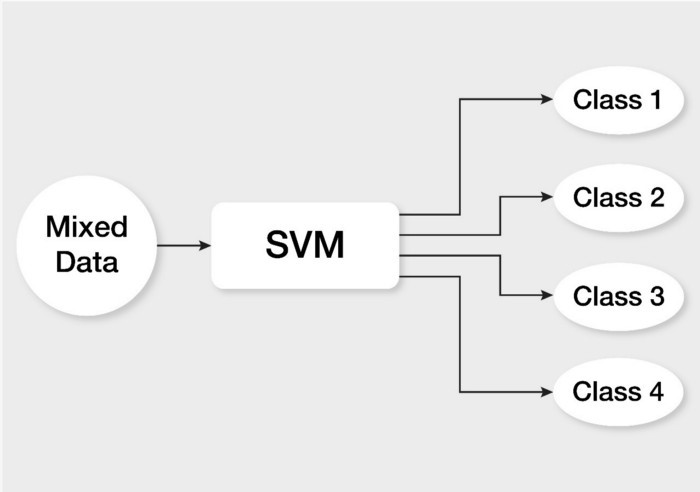}\caption{Representative diagram of Support Vector Machine}
\end{figure}

A Support Vector Machine (SVM) is a discriminative classifier formally defined by a separating hyperplane. In other words, given labeled training data (supervised learning), the algorithm outputs an optimal hyperplane that categorizes new examples. In two-dimensional space, this hyperplane is a line dividing a plane into two parts wherein each class lay on either side.
At first, approximation what SVMs do is to find a separating line (or hyperplane) between data of two classes. SVM is an algorithm that takes the data as an input and outputs a line that separates those classes if possible.
Let’s begin with a problem. Suppose we have a dataset as shown below and we need to classify the red rectangles from the blue ellipses (let’s say positives from the negatives). So our task is to find an ideal line that separates this dataset into two classes (say red and blue).

\begin{figure}[h]
\centering
\includegraphics[width=0.7\columnwidth]{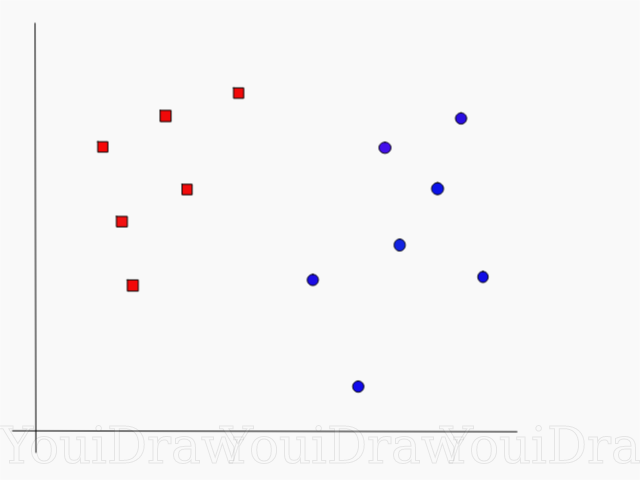}
\end{figure}
But there isn’t a unique line that does the job. We have infinite lines that can separate these two classes. So, the question is, how SVM finds the ideal one.
It’s visually quite intuitive in this case that the yellow line classifies better. But, we need something concrete to fix our line. The green line in the image above is quite close to the red class. Though it classifies the current datasets it is not a generalized line and in machine learning, our goal is to get a more generalized separator.

\begin{figure}[h]
\centering
\includegraphics[width=0.7\columnwidth]{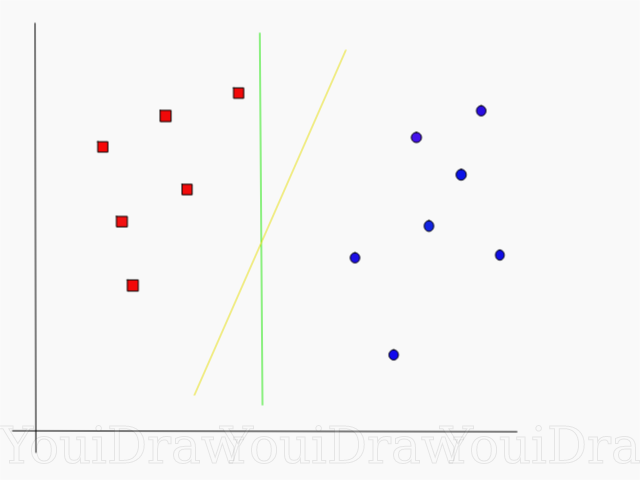}
\end{figure}
\begin{figure}[h]
\centering
\includegraphics[width=0.7\columnwidth]{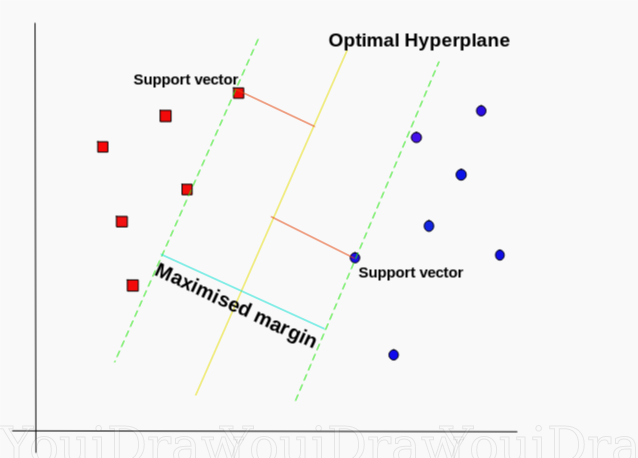}\caption{Binary classification using Support Vector Machine (SVM)}
\end{figure}
According to the SVM algorithm, we find the points closest to the line from both the classes. These points are called support vectors. Now, we compute the distance between the line and the support vectors. This distance is called the margin. Our goal is to maximize the margin. A hyperplane in an n-dimensional Euclidean space is a flat, n-1 dimensional subset of that space that divides the space into two disconnected parts. The hyperplane for which the margin is maximum is the optimal hyperplane.
Thus SVM tries to make a decision boundary in such a way that the separation between the two classes is as wide as possible.

\subsubsection{\textbf{Artificial Neural Network (ANN)}}
Artificial neural networks (ANNs), usually simply called neural networks (NNs), are computing systems vaguely inspired by the biological neural networks that constitute animal brains.
An ANN is based on a collection of connected units or nodes called artificial neurons, which loosely model the neurons in a biological brain. Each connection, like the synapses in a biological brain, can transmit a signal to other neurons. An artificial neuron that receives a signal then processes it and can signal neurons connected to it. The "signal" at a connection is a real number, and the output of each neuron is computed by some non-linear function of the sum of its inputs. The connections are called edges. Neurons and edges typically have a weight that adjusts as learning proceeds. The weight increases or decreases the strength of the signal at a connection. Neurons may have a threshold such that a signal is sent only if the aggregate signal crosses that threshold. Typically, neurons are aggregated into layers. Different layers may perform different transformations on their inputs. Signals travel from the first layer (the input layer) to the last layer (the output layer), possibly after traversing the layers multiple times.
But ANNs are less motivated by biological neural systems, there are many complexities to biological neural systems that are not modeled by ANNs.
\begin{figure}[h]
\centering
\includegraphics[width=\columnwidth]{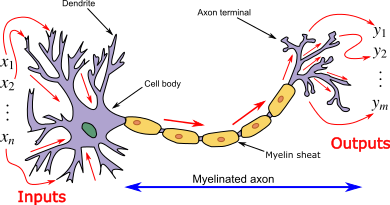}\caption{Representative diagram of Artificial Neural Network (ANN), inspired by neuron cells}
\end{figure}

\textbf{Characteristics of Artificial Neural Network:}

\begin{itemize}
\item It is a neurally implemented, mathematical model.

\item It contains a huge number of interconnected processing elements called neurons to do all operations.

\item Information stored in the neurons are the weighted linkage of neurons

\item The input signals arrive at the processing elements through connections and connecting weights.

\item It can learn, recall, and generalize from the given data by suitable assignment and adjustment of weights.

\item The collective behavior of the neurons describes its computational power and no single neuron carries specific information.

\end{itemize}

\textbf{Advantages of ANN}
\begin{itemize}
\item Problem in ANNs can have instances that are represented by many attribute-value pairs.
\item ANNs used for problems having the target function output may be discrete-valued, real-valued, or a vector of several real- or discrete-valued attributes.
\item ANN learning methods are quite robust to noise in the training data. The training examples may contain errors, which do not affect the final output.
\item It is used generally used where the fast evaluation of the learned target function may be required.
\item ANNs can bear long training times depending on factors such as the number of weights in the network, the number of training examples considered, and the settings of various learning algorithm parameters.
\end{itemize}

\subsubsection{\textbf{Single Layer Perceptrons}}
Input is multi-dimensional (i.e. input can be a vector): input x = ( I1, I2, .., In).
Input nodes (or units) are connected (typically fully) to a node (or multiple nodes) in the next layer. A node in the next layer takes a weighted sum of all its inputs.
\textbf{The rule:} The output node has a “threshold” t.

If summed input $>$ t, then it “fires” : (output y = 1).
Else (summed input $<$ t) it doesn't fire: (output y = 0).

\begin{figure}[h]
\centering
\includegraphics[width=\columnwidth]{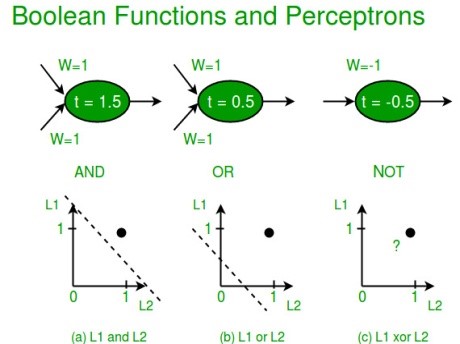}\caption{Boolean representations of perceptrons}
\end{figure}

\textbf{Limitations of Perceptrons}
\begin{enumerate}
\item The output values of a perceptron can take on only one of two values (0 or 1) due to the hard-limit transfer function.
\item Perceptrons can only classify linearly separable sets of vectors. If a straight line or a plane can be drawn to separate the input vectors into their correct categories, the input vectors are linearly separable. If the vectors are not linearly separable, learning will never reach a point where all vectors are classified properly.
\end{enumerate}

The Boolean function XOR is not linearly separable (Its positive and negative instances cannot be separated by a line or hyperplane). Hence a single layer perceptron can never compute the XOR function. This is a big drawback that once resulted in the stagnation of the field of neural networks. But this has been solved by multi-layer.

\subsubsection{\textbf{Multi-layer Perceptrons (MLP)}}
A Multi-Layer Perceptron (MLP) or Multi-Layer Neural Network contains one or more hidden layers (apart from one input and one output layer). While a single layer perceptron can only learn linear functions, a multi-layer perceptron can also learn non-linear functions.
\begin{figure}[h]
\centering
\includegraphics[width=0.8\columnwidth,height=60mm]{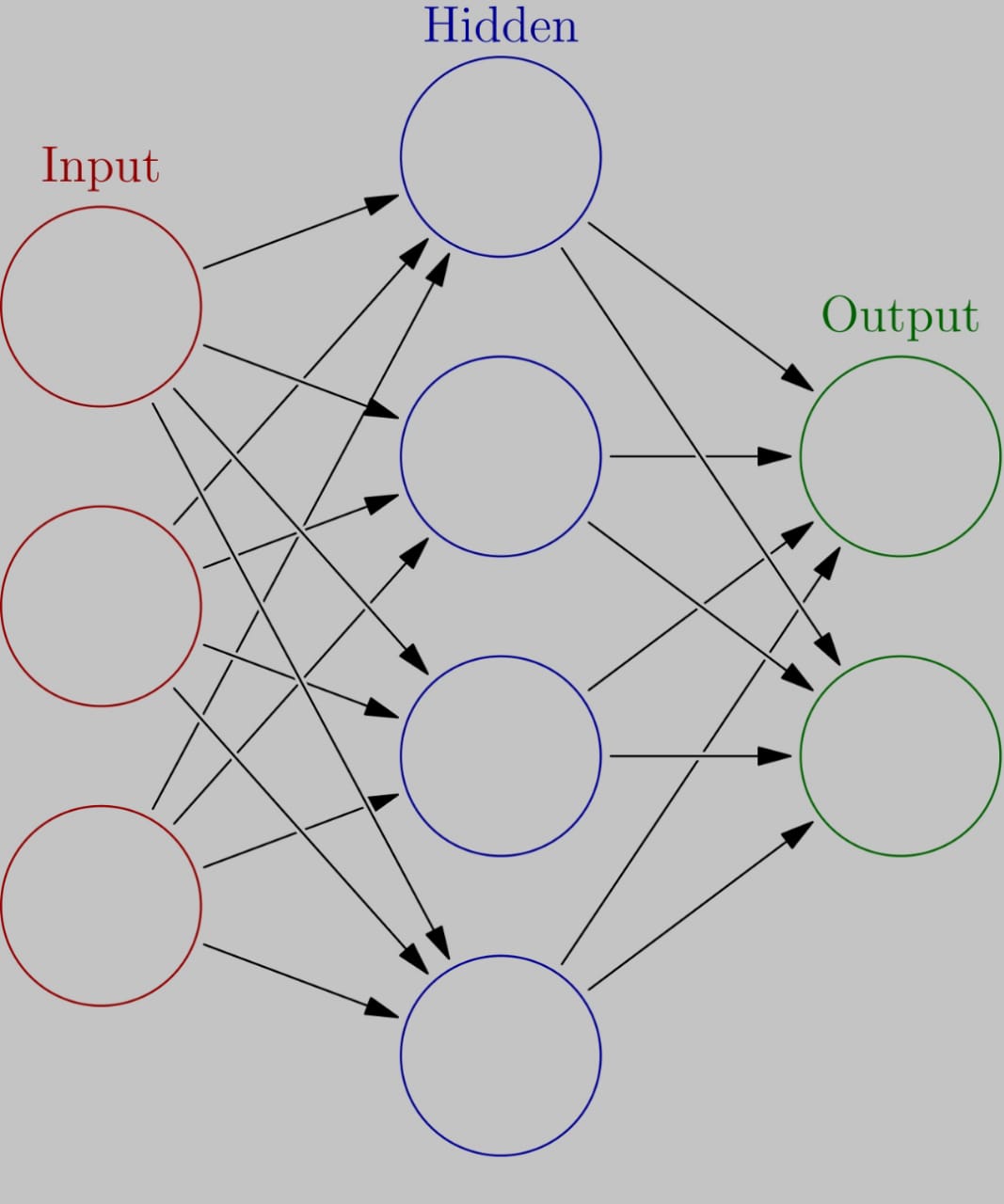}\caption{Simplified architecture of Multi-Layer Perceptron (MLP) with only one hidden layer.}
\end{figure}

This neuron takes as input x1,x2,….,x3 (and a +1 bias term), and outputs f(summed inputs+bias), where f(.) called the activation function. The main function of Bias is to provide every node with a trainable constant value (in addition to the normal inputs that the node receives). Every activation function (or non-linearity) takes a single number and performs a certain fixed mathematical operation on it. There are several activation functions, such as:

\textbf{Sigmoid: } takes real-valued input and squashes it to range between 0 and 1.
\begin{equation}
\sigma(x)=\frac{1}{1+e^{-x}}
\end{equation}

\textbf{tanh: } takes real-valued input and squashes it to the range [-1, 1 ].
\begin{equation}
tanh(x)=2\sigma (2x)-1
\end{equation} 

\textbf{ReLU: } ReLu stands for Rectified Linear Units. It takes real-valued input and thresholds it to 0 (replaces negative values to 0 ).
\begin{equation}
f(x)=max(0,x)
\end{equation}

\textbf{Leaky ReLu: }has a small positive slope in its negative area, enabling it to process zero or negative values.

\textbf{Parametric ReLu: } allows the negative slope to be learned, performing backpropagation to learn the most effective slope for zero and negative input values.

\textbf{Softmax: } normalizes outputs for each class between 0 and 1 and returns the probability that the input belongs to a specific class.

\textbf{Swish: } performs better than ReLu with a similar level of computational efficiency.
\subsubsection{\textbf{Working Principle of Simple Neuron:}}
Let there are two neurons X and Y which is transmitting a signal to another neuron Z. Then, X and Y are input neurons for transmitting signals and Z is the output neuron for receiving a signal. The input neurons are connected to the output neuron, over interconnection links (A and B) as shown in the figure:

\begin{figure}[h]
\centering
\includegraphics[width=\columnwidth]{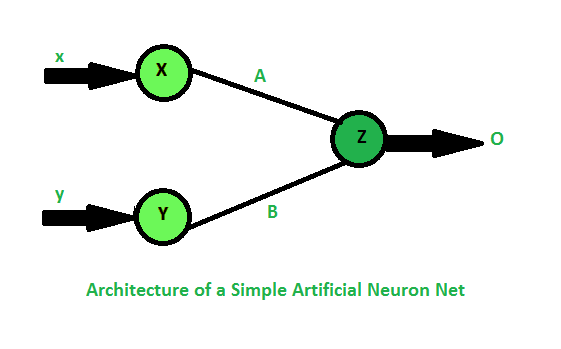}\caption{Working of a simple ANN}
\end{figure}

For above neuron architecture , the net input has to be calculated in the way .
\begin{equation}
I = xA + yB
\end{equation}

where x and y are the activations of the input neurons X and Y . The output z of the output neuron Z can be obtained by applying activations over the net input .
\begin{equation}
Output = Function ( \text{net input calculated} )
\end{equation}

The function to be applied over the net input is called an activation function. There is various activation function possible for this.

\section{Results and discussions}
We are given a non-manipulated dataset of 10 individuals. Each of them is having 10 different types of finger movements which include individual and combined movements. The movements are classified as follows-
\begin{enumerate}
\item Thumb (T)
\item Index (I)
\item Middle (M)
\item Ring (R)
\item Little (L)
\item Thumb-Index (T-I)
\item Thumb-Middle (T-M)
\item Thumb-Ring (T-R)
\item Thumb-Little (T-L)
\item Hand close (HC)
\end{enumerate}
We have to increase the classification accuracy as much as possible (preferably $>$95\%) where the classification is based on the 10 classes of finger movements as described above. 

\subsection{Optimization of dimension reduction}
While employing PCA for classification, feature reduction becomes evident as some of the features are waived off thereby limiting most of the information in the chosen features for the entire dataset. This feature reduction technique, once employed, also leads to a shift in the accuracy of classification as it is up to the programmer who writes the code and uses feature reduction as is necessary. Therefore, to calculate the classification accuracy, we have changed the select components after employing PCA and used the ANN classifier every time.

Our goal is to keep the maximum amount of information present in the feature set and simultaneously reduce the number of principal components to reduce the computational cost. Following is the graph of several selected components against the variance.

\begin{figure}[h]
\centering
\includegraphics[width=\columnwidth]{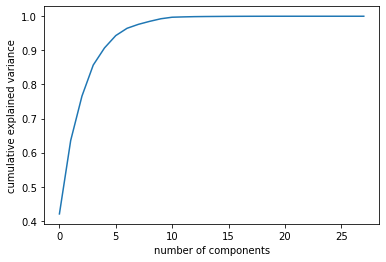}\caption{Representative graph of cumulative explained variance against number of components selected by PCA.}
\end{figure}

Thus by keeping the number of components $>$ 10, we can keep almost all the information of the dataset. This is also evident from the following graph of accuracy plot against the number of components selected both for PCA and LDA. It is evident from the graph below that PCA is more evident and robust as compared to LDA for feature selection and hence classification accuracy is slightly lesser in the case of LDA as compared to PCA. 

\begin{figure*}[h]
\centering 
\subfigure[LDA+ANN non-normalized]{\includegraphics[width=0.8\columnwidth]{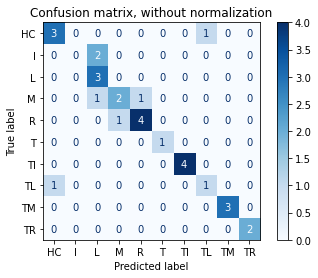}}
\subfigure[LDA+ANN normalized]{\includegraphics[width=0.8\columnwidth]{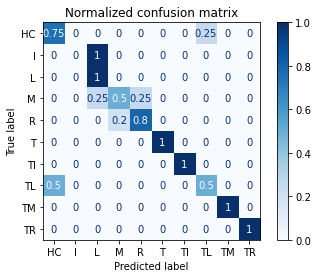}}
\subfigure[PCA+ANN non-normalized]{\includegraphics[width=0.8\columnwidth]{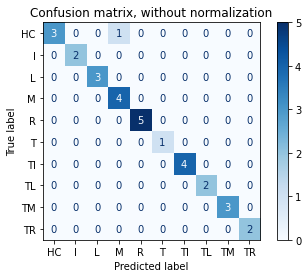}}
\subfigure[PCA+ANN normalized ]{\includegraphics[width=0.8\columnwidth]{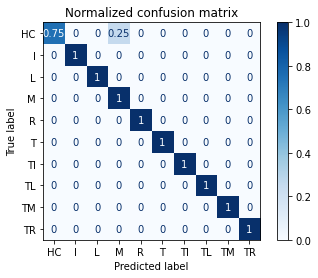}}
\caption{Confusion matrix for ANN classifier with PCA and LDA}
\end{figure*}

\begin{figure}[h]
\centering
\includegraphics[width=\columnwidth]{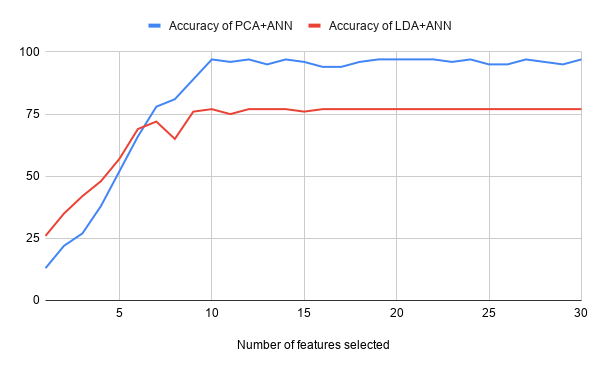}\caption{Plot of accuracy of ANN classifier withPCA and LDA and different numbers of features selected.}
\end{figure}

\begin{figure*}[h]
\centering 
\subfigure[LDA+SVM non-normalized]{\includegraphics[width=0.8\columnwidth]{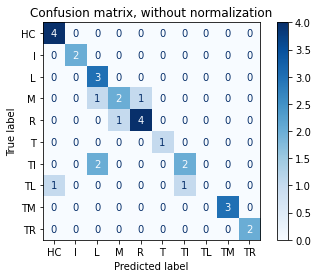}}
\subfigure[LDA+SVM normalized]{\includegraphics[width=0.8\columnwidth]{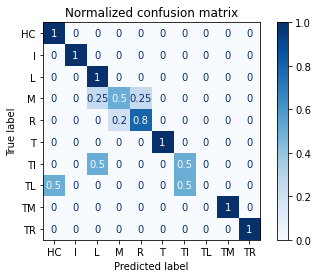}}
\subfigure[PCA+SVM non-normalized]{\includegraphics[width=0.8\columnwidth]{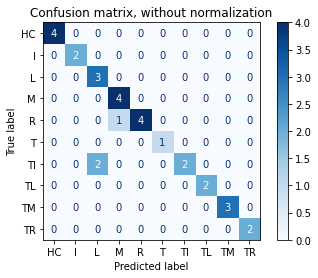}}
\subfigure[PCA+SVM normalized ]{\includegraphics[width=0.8\columnwidth]{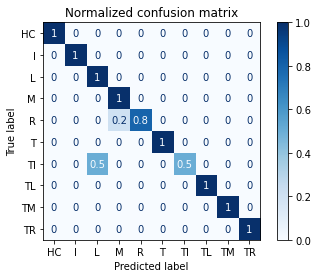}}
\caption{Confusion matrix for SVM classifier with PCA and LDA}
\end{figure*}

\subsection{Optimization of train-data and test-data sizes}
For a given dataset, dividing it into proper training and testing sets plays a very pivotal role in any application-oriented machine learning step. If the number of elements in training and testing sets is not in the proper ratio, it may lead the system to under-learn or over-learn, both of which are unexpected to avoid improperly learning of the algorithm by the system.
Hence, based on the type of dataset, there exists an optimum division of data for training and testing respectively.
For our data size of 480, we have taken 5 cases of testing and training sets, and noted the accuracy in each of the following cases below. 
\begin{table}[h]
\centering
\caption{Comparison of different train-split size with their corresponding accuracy value. The bold digits represent the optimum result. The experimentation is done solely on PCA+ANN comnbination.}
\label{tab:my-table}
\begin{tabular}{@{}lll@{}}
\toprule
\textbf{TRAINING SET} & \textbf{TESTING SET} & \textbf{\begin{tabular}[c]{@{}l@{}}ACCURACY\\ (in \%)\end{tabular}} \\ \midrule
350 & 130 & 78.67 \\
375 & 105 & 81.25 \\
400 & 80 & 89.50 \\
425 & 55 & 92.90 \\
\textbf{450} & \textbf{30} & \textbf{96.67} \\ \bottomrule
\end{tabular}
\end{table}

For feature reduction in each case, we have used Principal Component Analysis (PCA) and for classification, Artificial Neural Network or the ANN classifier is used. By slight observation of the testing and training set table, we observe as the data in the training set increases, the accuracy increases almost at a linear rate. The accuracy is maximum ( 96.67 \%) when the data size of the training set is 450. Hence, this proportion would be chosen for all the classifications and to form the confusion matrix of the same. 

\subsection{Accuracy and Confusion matrix}
After receiving the data and performing the cleaning, pre-processing, and wrangling, we feed it to a model to get the output as a probability. But there is a drawback in this procedure- we don’t know the effectiveness of our model. We need the most optimized performance and the performance is directly proportional to the effectiveness and we need a tool to measure the effectiveness. Hence the need for a confusion matrix comes into play.
\subsubsection{\textbf{Accuracy}}
As we can infer from the name itself, accuracy is the measure of correctness of the classifier after some training. It depends on various factors like the size of the training dataset, type of dataset, seeding value, type of classifier used, etc. 
When our entire dataset was taken as features, the classification accuracy fell below 20\%. Therefore, to cater to that, we have prepared our algorithm with 16 features along with their moments. From the table below, it can also be observed that among the two classifiers, ANN seems to give the highest possible accuracy.

\begin{table}[]
\centering
\caption{Classification accuracy of SVM and ANN classifiers with PCA and LDA implemented, with and without feature extraction}
\label{tab:my-table}
\begin{tabular}{|c|c|c|c|c|c|}
\hline
\multirow{2}{*}{Channel} &
  \multirow{2}{*}{\begin{tabular}[c]{@{}l@{}}Feature\\ Selection\end{tabular}} &
  \multicolumn{2}{c|}{\begin{tabular}[c]{@{}l@{}}With \\ features\end{tabular}} &
  \multicolumn{2}{c|}{\begin{tabular}[c]{@{}l@{}}Without \\ features\end{tabular}} \\ \cline{3-6} 
                                                                                       &     & SVM     & ANN     & SVM     & ANN     \\ \hline
\multirow{2}{*}{\begin{tabular}[c]{@{}l@{}}Channel \\ A\end{tabular}}                  & PCA & 88.25\% & 93.67\% & 13.58\% & 14.22\% \\ \cline{2-6} 
                                                                                       & LDA & 72.20\% & 73.32\% & 11.00\% & 12.43\% \\ \hline
\multirow{2}{*}{\begin{tabular}[c]{@{}l@{}}Channel \\ B\end{tabular}}                  & PCA & 89.54\% & 94.50\% & 15.92\% & 16.20\% \\ \cline{2-6} 
                                                                                       & LDA & 75.00\% & 76.33\% & 11.00\% & 13.85\% \\ \hline
\multirow{2}{*}{\begin{tabular}[c]{@{}l@{}}Channel \\ A and B\end{tabular}} & PCA & 90.00\% & 96.67\% & 17.11\% & 18.25\% \\ \cline{2-6} 
                                                                                       & LDA & 76.67\% & 76.67\% & 11.67\% & 15.33\% \\ \hline
\end{tabular}
\end{table}
\subsubsection{\textbf{Confusion matrix}}
As we described briefly the need for confusion matrix, is a performance measurement technique for machine learning classification. From its outlook, it is similar to an NxN square matrix, N is the number of classes of data classification. The rows and columns of this matrix represent the number of actual data classes and predicted classes respectively. Therefore, ideally, the matrix should be a diagonal matrix for any classifier with all the elements being 1 for a 100\% accurate prediction. But in practice, not all the predictions are always correct and this results in the matrix also showing the positions and number of incorrect classifications.
From the table above, four types of classification are done- PCA-SVM, PCA-ANN, LDA-SVM, and LDA—ANN. For each type, we can have a confusion matrix. Hence, we have 4 types of confusion matrix with 10 classes each.

As from the figures above, the confusion matrix helps us in getting an idea of the errors and their types as done by the classifier. The color of the boxes shows the accuracy of classification, black being the highest possible accuracy. The name confusion matrix is aptly justified as we can see the confusion of a classifier while making a prediction.

\end{document}